# Phononic and electronic features of diamond nanowires

Farah Marsusi[*1,], Seyed Mostafa Monavari

[1]Department of Physics and Energy Engineering, Amirkabir University of Technology, PO Box 15875-4413, Tehran, Iran

**Abstract**

By using density functional theory, we have studied vibrational features, Raman activities, and electronic properties of ultrathin hydrogen-passivated diamond nanowires (H-DNWs). Confinement imposes the softening of transverse acoustic modes and the hardening of the low-optical modes. Higher frequencies optical modes are more influenced by the surface termination effect. Surface effect splits transverse (TO) and longitudinal (LO) optical modes with a redshift from the bulk diamond. Electronic inspection shows that the surface terminated hydrogen atoms have the main contribution to the conduction bands of H-DNWs. This causes, contrary to our expectations from the quantum confinement effect, the electronic bandgap of H-DNWs becomes smaller than bulk diamond.

1. Introduction

Diamond shows a combination of outstanding properties including wide band gap, mechanical strength, negative electron affinity, high refractive index, transmissivity, high heat conductivity, and tunable electronic properties. Most of these features arise from the strong $sp^3$ C-C covalent bonding in the diamond lattice structure. The integration of diamond superior properties and nanoscale features gives exclusive properties to diamond nanostructures. Amongst them, one-dimensional diamond nanowires (DNWs) are still in the early stages of a controllable synthesis [1]. However, simulations and experiments have verified that DNWs have a number of potential applications, for example, in nanomechanical and optoelectronic nanodevices. Depending on the type of materials used and the shape and orientation of DNWs, a different number of synthesis routes have been applied by the experimental works [2]. For example, DNWs in 60-90 nm diameter with high crystallinity and structural uniformity were synthesized utilizing chemical vapor deposition (CVD) at atmospheric pressure [3]. The process was assisted by hydrogen flow over a Fe catalyst solution dispersed on a silicon substrate. Hydrogen commonly is applied in the diamond growth process and enhances diamond nucleation as well as facilitates transforming the carbon sp- and $sp^2$-bonds into $sp^3$-hybridized bonds [1,4]. In another work, using microwave plasma CVD process $sp^2/sp^3$ carbon ratio was controlled by incorporation of $N_2$ gas and partially oriented DNWs at 5 nm diameter, covered with $sp^2$-bonded, were formed [5].

---

[1] *Corresponding author. E-mail: marsusi@aut.ac.ir

Although significant progress in recent years has been made in the successful synthesis of DNWs, a reproducible synthesis remains a challenge. [4]. Once the reproducible synthesis is understood, DNWs are expected to be valuable in the diversity of applications such as ultra-low threshold electron field emissions devices with high-brightness [1, 6], high-performance nano-electromechanical switches, and biosensors [7-8]. Therefore, it requires more studies to understand the underlying physics of thermal and electronic features of DNWs. Still, we need to know the role of surface capping, since it may be a tool to control DNWs properties.

In a molecular dynamics (MD) study, very low thermal conductivity coefficients (κ) of 2-30 $Wm^{-1} k^{-1}$ were predicted at 300 K for ultrathin DNWs that their diameters are between 0.4-2 nm, respectively [9]. According to the MD prediction, the highest thermal κ belongs to those wires that their growth axis is in [110] direction of the bulk parent diamond and the phonon group velocities were inferred responsible for the anisotropic thermal conductivities. In another study based on the *ab initio* Peierls-Boltzmann phonon transport equation (PBTE), the behavior of κ in terms of size and temperatures was studied for larger wires with diameters between 10 nm and 1000 mm [10]. They found that the highest value of κ belongs to those wires that their axes are in [100] direction. The value of κ at room temperature was found 60 $Wm^{-1} k^{-1}$ for the smallest 10 nm wire and it approaches about 50% of the bulk diamond at 670 nm wire [10]. According to PBTE outcomes, κ is more strongly affected by the temperatures in DNWs than silicon nanowires (SiNWs). This latter interesting property can be used to adjust thermal conductivity by the size of the wire.

Raman spectroscopy has been extensively employed for exploring the phonon band structure of nanostructures. Micro-Raman spectrum of a DNW with a diameter between 60 and 90 nm growth in a carbon nanotube (CNT) shows a signature peak of the diamond structure at 1332 $cm^{-1}$[1]. In a theoretical study, based on a local bond-polarization model and the Born potential, the confined optical phonons and Raman scattering of diamond nanowires have been studied and it was found that the Raman peaks shifts to lower energies in DNWs with diameter less than 2 nm [11]. However, still, data on vibrational features and shifts versus the size of DNWs are not available. Yet, the confined phonon modes in the low-frequency regime and the surface effect at the high-frequency window are not clearly understood. While these modes are expected to give important structural information. Yang et al. have shown that the highest-frequency optical modes of SiNWs are split with a size-dependent redshift resulting from surface effect. The observed redshifts are used to estimate the size of SiNWs [12]. Despite the same lattice structures, diamond crystal shows unique physical properties different from silicon. Since there is no *p* electron in the core of the carbon atom and the valence 2*p* electrons are very close to the nuclei. Therefore, we need to know the impact of terminating the $sp^3$ strong network by the surface at DNWs.

Based on the density functional perturbation theory (DFPT), we investigate the phonon dispersion structure and Raman spectra of ultrathin hydrogen-passivated DNWs at between 0.73-1.2 nm diameters. We observe emerging of some confined vibrational modes in lower optical frequencies that carry size-dependent information and influence the heat capacity $C_V$ in lower temperatures. On the other hand, our results reveal the splitting of the highest-frequency optical TO and LO modes with redshifts from the related value in bulk diamond.

In addition to thermal behaviors, DNWs exhibit unique optoelectronic properties that are valuable for the diversity of important applications. Previous theoretical studies of ultrathin DNWs with diameters of about 1-2 nm showed that the bandgap of these wires depends on the combination of many factors such as size, morphology, and surface chemistry [13]. In another work, the effect of various dopants on the stability of ultrathin cylindrical DNWs was studied [2,14].

In our study, we focus on the DNWs with electronic closed-shell structures. Our results show that the closed-shell fully hydrogenated DNWs are non-magnetic materials. However, defects can induce magnetism in carbon allotropes [15]. We found that the local atomic magnetic moment may be induced in the DNWs if there is a free dangling bond on the surface that is not passivated by for example a hydrogen atom. Since hydrogen flow often be used during the CVD synthesis process, we stabilize the dangling bonds on the surface by hydrogen atoms. Our results show that hydrogen atoms on the surface of DNWs have almost the same impact as they have on diamondoids (hydrogen-terminated diamond-like nanoparticles). Diamond nanostructures and diamondoids show high selectivity in binding to chemical groups and biological materials [16]. This property gives particular applications in bio-sensing research and technology. Today, highly sensitive and selective biosensors are of great urgency for operating in multiple applications including disease diagnosis and environmental monitoring. DNWs can be one of the promising candidates due to their biocompatibility as well as good electrochemical properties. For sensing applications, the biosensor surface must first successfully interact with a single analyte ion or molecule in a way that it changes some measurable quantities like conductivity or fluorescence intensity. One way to improve the binding between biosensor and analyte species is functionalizing surfaces with amine-based chemical groups. Here we explore the possibility of $NH_2$-functionalized DNW, since amine-functionalized surfaces can make covalent bonding to biological systems such as DNA and carbohydrates. [17-18]. Our results show a significant possibility of functionalized $NH_2$ DNW via notable binding energy of about 4 eV.

The rest of this paper is organized as follows: in section 2 we explain in detail our computational methods. In subsections 3.1 and 3.2 we present and interpret phonon behaviors and Raman spectra of bulk diamond and DNWs, respectively. The electronic properties are presented in subsection 3.3. Finally, a summary and our conclusions are given in section 4.

## 2. Computational details

The repeating pattern of atoms in the nanowires is along the wire axis. The [uwv] standard orientation of the parent crystal along the wire axis is called the wire orientation. The axes of DNWs in this work align along [110] direction with diameters of 0.73 and 1.2 nm (By including hydrogen-terminated atoms). They are constructed in cylindrical morphology. First-order Raman scattering of solid crystals arises from a single phonon at Γ-point. Previous calculated vibrational modes of the bulk diamond by using local density approximation (LDA) are very close to the experiments, in particular those modes that happen at the Γ-point [19]. This encouraged us to perform our calculations by using LDA functional. We used the same Norm-conserving pseudopotentials based on von Barth and Car scheme [20], as used in Ref. [19].

A tetragonal supercell with the lengths of $L_x=L_y=20$ Å was used for DNWs to guarantee a vanishing interaction between the periodic images along x- and y-directions. Then, all atomic positions and lattice vectors along the z-direction are fully relaxed until the atomic forces become less than 1 meV/Å. Thermal quantities of DNWs are compared with the corresponding values obtained at the same level of the theory from the bulk diamond calculations. Lattice constant and frequencies of bulk diamond with the given pseudopotential were converged at large kinetic cutoff energy of 65 Ry. We used a 16×16×16 and 1×1×16 Monkhorst-Pack **k**-mesh for bulk diamond and DNWs, respectively [21]. We applied the $10^{-12}$ a.u and $10^{-15}$ a.u. to converge energy difference in the ground state and in the response function calculations of the phonon modes, respectively. All calculations are done by using the Quantum Espresso code (QE) [22]. Phonon modes and orbital frontiers are visualized by using the XCrySDen and Vesta software, respectively [23-24].

## 3. Results and discussions

### 3.1. Bulk diamond

To understand the phononic features and Raman active vibrational modes of DNWs we first need to analyze the vibrational structures of the bulk diamond. We present the diamond LDA phonon dispersion curves and the corresponding vibrational density of states (VDOS) along the symmetry paths in Figure 1. The agreement with the previous data supports the accuracy of our calculations [19, 25]. According to this figure, unlike most of the cubic semiconductors with a motif of two tetrahedrally bonded atoms, diamond TA branches do not show flat features as it is reflected in the smooth VDOS in the corresponding energy window. In contrast to Si and Ge crystals [26], and in agreement with Refs. [25] and [19], the maximum of VDOS does not meet the Γ-point. This was previously understood by a large interatomic force constant between the reference carbon atom and its second-nearest neighbors in diamond [19]. Our results show that the three degenerate LDA optical modes at the center of the first Brillouin zone (BZ) happen at 1348 cm$^{-1}$, 24 cm$^{-1}$ above the experimental value of 1324 cm$^{-1}$ [27-28]. These three degenerate optical modes stem from the opposite moving of the two carbon atoms in the primitive cell in the three orthogonal directions. There is a small difference between our outcomes and LDA-predicted optical frequencies in Ref. [19]. We found that this slight discrepancy originated from the smaller cutoff energy (55 Ry) that was applied in Ref. [19]. When we used the cutoff energy of 55 Ry, we obtained the same optimized lattice constant and therefore phonon spectrum as in the Ref. [19].

In Table 1, we summarize our results at high symmetry points and compare them with the previous theoretical and experimental data. The theoretical modes are influenced by the lattice parameter and in general LDA-modes are harder than GGA. The origin of Raman peaks is the variation of the electronic polarizability. According to the Raman selection rules, this may include those vibrational modes that their symmetric representations are in the quadratic form of $x^2$, $y^2$, $z^2$, xy, xz, and yz terms. Bulk diamond belongs to the $O_h$ (m-3m) point group. The three-degenerate optical modes at G-point with the energy of 1332 cm$^{-1}$ have $T_{2g}$ ($G_{25'}$) symmetry and are Raman active. The basis functions associated with this symmetry are quadratic terms of xz, yz, xy [29], and Raman active. With a small blue shift, our LDA Raman peak is at 1348 cm$^{-1}$.

Figure 1. Phonon band structure and vibrational density of states (VDOS) of the bulk diamond.

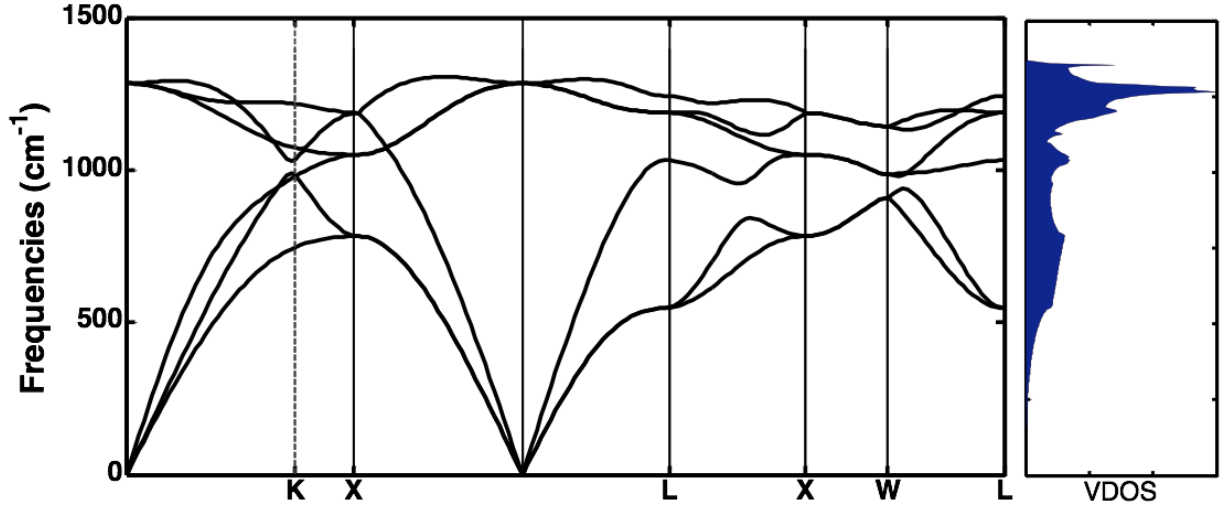

**Table 1**: Lattice parameter in Bohr and phonon frequencies of diamond crystal at the high-symmetry points Γ, X and L, in cm$^{-1}$. Data from previous theoretical and experimental works are given for comparison.

|       | $a_0$  | ΓO   | $X_{TA}$ | $X_{TO}$ | $X_{LO}$ | $L_{TA}$ | $L_{LA}$ | $L_{TO}$ | $L_{LO}$ |
|-------|--------|------|----------|----------|----------|----------|----------|----------|----------|
| LDA   | 6.640  | 1348 | 794      | 1123     | 1239     | 554      | 1098     | 1256     | 1282     |
| LDA[a]| 6.67   | 1324 | 800      | 1094     | 1228     | 561      | 1080     | 1231     | 1275     |
| GGA[b]| 6.743  | 1289 | 783      | 1057     | 1192     | 548      | 1040     | 1193     | 1246     |
| Exp.[c]| 6.740 | 1332 | 807      | 1072     | 1184     | 550      | 1029     | 1206     | 1234     |

From Ref. [19]
From Ref. [25]
From Ref. [27]

*3.2. Phononic features of diamond nanowires*

The smallest 0.73 nm wire considered in this study, consists of 28 atoms. With increasing the size of the wire to 1.2 nm, this number increases to 62 atoms, as shown in Figures 2(a). We found no negative frequency, which means these DNWs are stable. Phonon dispersion curves of 1.2 nm wire are shown in Figure 3(a). Four acoustic branches are observed in the *first* region: three branches stem from the 3D translational degree of freedom and one extra branch, which is absent in bulk diamond, is associated with the rotation about the wire axis and stems from confinement in x- and

y- directions. Longitudinal acoustic branch (LA stretching modes), with the displacement of atoms along the wire axis, and rotational branch (RA) is almost linear in $q_z$ *around* the BZ center. The two other acoustic branches are softer transverse (TA) *bending* modes. These two modes are associated with the atomic in-plane displacements perpendicular to the wire axis. TA modes show nearly quadratic dispersion proportional to $q_z^2$ around the center of BZ. Figure 3(b) zooms in on the acoustic branches of 1.2 nm wire. The pictorial illustrations of atomic displacements on the BZ center are also shown in this figure. The group velocities of TA and LA of 1.2 nm wire are reduced by 66% and 2% from the bulk diamond, respectively. Therefore, DNWs should have smaller Young's modulus than the bulk diamond, as predicted previously by the MD simulations [9,30-31]. By increasing the size of the wire sound modes, especially TA, become harder. The lowest optical mode of 0.73 nm wire starts at 268 cm$^{-1}$ and a progressive softening of this mode is observed by increasing the size of the wire. We found an 80 cm$^{-1}$ redshift in the lowest optical frequency when the size of the wire increases to 1.2 nm (at 177 cm$^{-1}$). In Figure 2(b) VDOS of the 0.73 nm wire resulted from C-C stretching and C-H bending modes with energies below 1500 cm$^{-1}$ are compared with the VDOS of the bulk diamond. The frequencies above 1500 cm$^{-1}$ indicate C-H stretching modes and are not shown here, because they are not of any interest in the present study. Spatial confinement effect in two lateral directions characterizes the phonon spectra by a large number of branches and sharp peaks (van Hove singularities), compared with the bulk diamond. Another remarkable feature in Figure 2(b) is the development of new modes in the region less than about 500 cm$^{-1}$, which are absent in bulk diamond. The contributions of the surface hydrogen atoms and the carbon atoms are separately specified in this figure. Depending on the contribution of hydrogen atoms on the collective atomic motions, two distinct regions are noted: (i) low-frequency region between about 0-800 cm$^{-1}$ with negligible surface contributions, (ii) high-frequency region above 800 cm$^{-1}$ with significant contributions of hydrogen atoms.

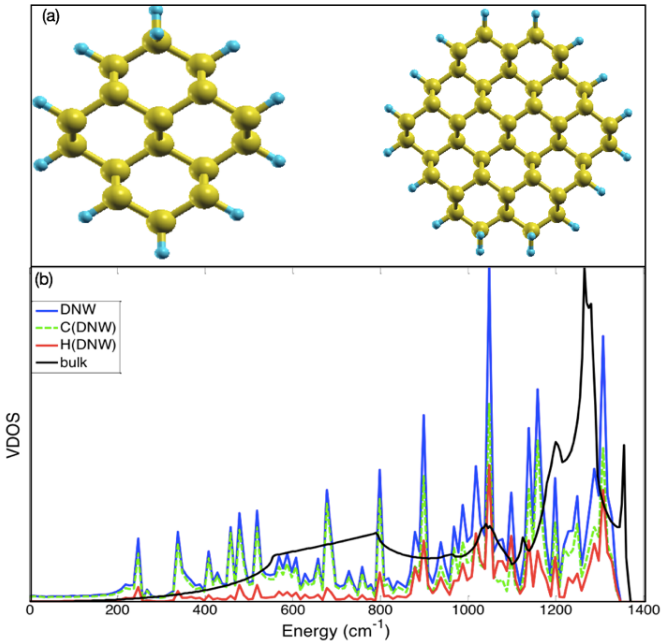

Figure 2. (a) The relaxed structures of 0.73 nm H-DNWs consists of 16 C and 8 H atoms and the 1.2 nm wire consists of 42 C and 20 H atoms. (b) VDOS of bulk (black-solid line) and 0.73 nm H-DNW (blue-solid line). VDOS decomposed into contributions of carbon (in green-dashed line) and hydrogen atoms (in red-line) are also shown in this figure. At high-frequency optical modes, DNW shows a clear red-shift from the bulk diamond.

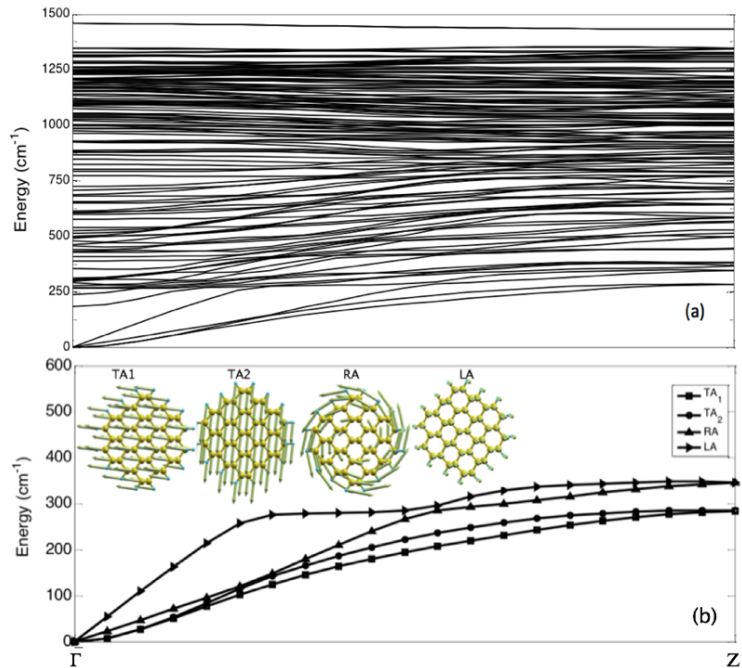

Figure 3. (a) Phonon dispersion curve of the 1.2 nm DNW. Frequencies above 1500 cm$^{-1}$, which are attributed to the C-H bending and stretching modes, respectively, are not shown. Point Z shows the boundary point of the first BZ. (b) A zoom illustration of the small frequencies acoustic dispersion lines of the panel (a). The corresponding eigenmodes at Γ-point are also shown.

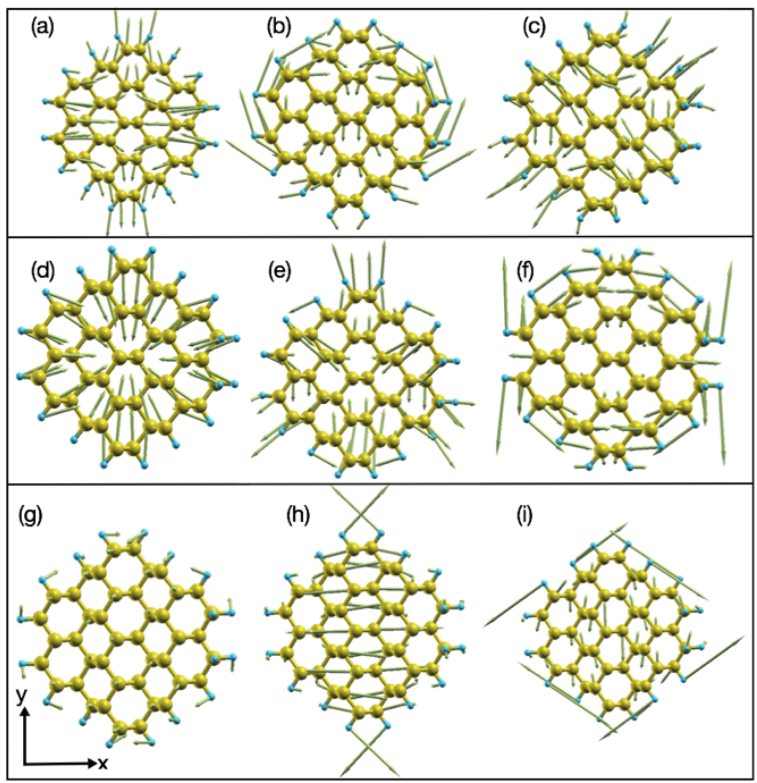

Figure 4. (a)-(f) Six confined modes of a 1.2 nm H-DNW with energy less than 400 cm$^{-1}$. RBM mode is shown in the panel (d), and LO mode in panel (g) with atomic vibrations along the z-direction. The two highest-frequency TO optical modes are shown in panels (h) and (i).

Since surface hydrogen atoms have a relatively lower impact on the first region the new peaks developed in the first region, in particular those below 500 cm$^{-1}$, are the subsequence of confinement. Radial breathing modes (RBM) at 509 cm$^{-1}$ for 0.73 nm wire with a collective correlated movement of all atoms along the radial direction is an example of a confined mode. By a certain redshift, RBM mode happens at 318 cm$^{-1}$ for larger 1.2 nm, see Figure 4(d). Meanwhile, the larger wire is still very thin and the dynamics of atoms are influenced by the confinement. The contributions of carbon atoms to collective motion becomes more clear in the 1.2 nm wire and the Raman activity corresponding to RBM increases with the size of the wire through increasing the chance of more variation in the polarizability during atomic vibrations. Though, one expects the intensity gradually decays and finally ends up in the large wires. Some other interesting confined modes for 1.2 nm wire which happen below 400 cm$^{-1}$ and are shown in Figure 4. Like RBM, other confined modes also display redshift in energy with increasing the size of the wire. However, by the exception of the RBM mode, all other modes show a weak or no Raman signal due to small change in the polarizability or symmetric considerations, and consequently cannot be taken as a size indicator of wire in Raman spectroscopy.

In the *second* region, we are interested in the highest-frequency optical modes which produce three-degenerate peaks of the first-order Raman spectrum in the bulk diamond. These are the two highest-frequencies in-plane transverses (TO$_1$ and TO$_2$) modes with atomic displacements perpendicular to the wire axis and one longitudinal (LO) mode, which produces atomic displacement along the wire axis. DNWs under study have cylindrical morphology and belong to C$_{2h}$ point group. Characteristic table of C$_{2h}$ group includes A$_g$, B$_g$, A$_u$, and B$_u$ terms, which of those only vibrations with A$_g$ and B$_g$ symmetries are quadratic functions and Raman active. In general, the number and the intensity of A$_g$-symmetric peaks in H-DNWs are more than Raman signals with B$_g$-symmetry. Previous studies show that the three-degenerate highest-frequency optical modes in SiNWs are split due to breaking of translational symmetry with certain redshifts from the corresponding value in bulk silicon [12]. Competition between two factors may determine the type and magnitude of the frequency shifts in the optical modes of the nanostructures: (i) increasing atomic force constant near the surface that is resulted from bond length contraction; (ii) reduction of the coordination number due to surface termination [32]. Our results show that C-C bond length decreases slightly by an average of 2% from the center to the surface of the DNWs. However, the calculated C-C bond lengths are on average longer than C-C bond length in bulk diamond. Therefore, the atomic force constant cannot become larger than the bulk diamond due to bond contraction near the surface. On the other hand, owing to small mass, hydrogen atoms at the surface show small resistance against the motion of the carbon atoms near the surface. Surface breaks the diamond strong sp$^3$ network and terminates further interactions for surface carbon atoms. This reduces the effective coordination number of the carbon atoms near the surface and is the origin of the softening of the vibrational modes. Hence, we expect redshifts in the high-frequency optical modes of DNWs from the bulk values. VDOS in Figure 2(b) shows a clear redshift in optical modes of DNW from the bulk diamond.

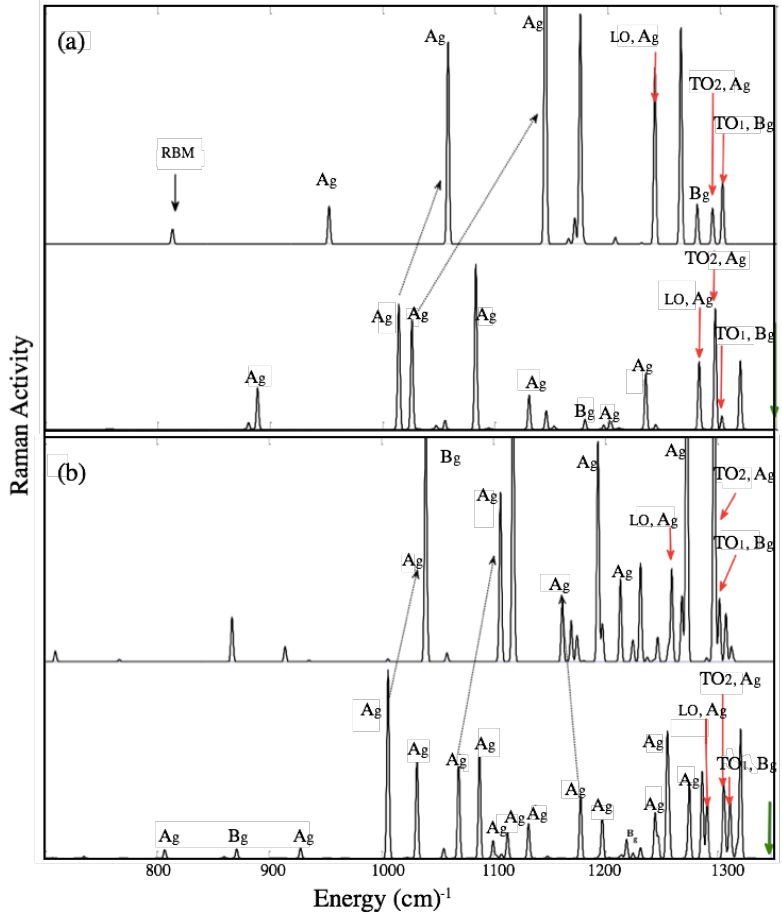

Figure 5. The Raman activities of (a) the 0.73 nm, and (b) 1.2 nm H-DNWs. For comparison, the Raman activities of DNWs when the mass of hydrogen atoms is artificially increased to 100 amu are shown at the top panel of each figure. The corresponding Raman line of the bulk diamond is shown by a green arrow at 1348 cm$^{-1}$. The symmetry of each line is labeled. Those modes with energy less than 700 cm$^{-1}$ has very small Raman activities and are not shown.

We present first-order LDA Raman spectra of DNWs in Figure 5 and specify the Raman peak of the bulk diamond by a green arrow for comparison. The corresponding symmetry terms of lines are labeled in this figure. The figures display clear redshifts from the bulk diamond and splitting the energy of the highest optical TO and LO modes. Contrary to RBM confined modes, the highest-energy optical modes show a softening trend in going from 1.2 nm to 0.73 nm. This can be explained by the fact that RBM modes originate from the confinement effect, while the high-frequency optical modes are influenced by the surface termination. Three sharp peaks with LO nature are observed in 0.73 nm wire at between 1000-1100 cm$^{-1}$ with the lowest mode at 1014 cm$^{-1}$. With increasing the size of the wire to 2.1 nm the number of sharp LO modes at this window is increased and four peaks are observed with the lowest mode at 1006 cm$^{-1}$, and hence these modes carry size information.

Owing to this fact that symmetric patterns along the x- and y- directions are not quietly the same, as seen in Figure 2(a), small splitting of about 10 cm$^{-1}$ is observed between the two-transverse TO$_1$ and TO$_2$ modes, see Table 2. The displacements of atoms by the highest-frequencies optical TO and LO modes of 1.2 nm wire are shown in Figure 4(g)-4(i). TO$_1$ modes in both wires show B$_g$ symmetry and smaller Raman activity. LO modes experience larger redshift and more energy (of about 30 cm$^{-1}$) is needed to excite TO modes, as concluded from Figure 5. The reason may be associated with the reduction of the C-C bond strength along the wire axes, since the lattice parameter "$a$" (along the wire axis) increases from the bulk value by 9% and 8% in the 0.73 nm and 1.2 nm wires, respectively, as seen in Table 2.

Table 2. lattice parameter "$a$" along the wire axis "z" in Bohr. The highest frequencies optical modes in three orthogonal directions x, y and z of H-DNWs in cm$^{-1}$. The star sign indicates the modes are calculated when the mass of hydrogen atoms is artificially increased to 100 amu.

| Size | $a$ | $TO_1$ | $TO_2$ | LO | $TO^*_1$ | $TO^*_2$ | $LO^*$ |
|---|---|---|---|---|---|---|---|
| 0.73 nm | 4.7947 | 1302 | 1295 | 1282 | 1302 | 1293 | 1242 |
| 1.2 nm | 4.7796 | 1313 | 1305 | 1290 | 1302 | 1297 | 1259 |

Mixing of C-C stretching modes with C-H bending modes can be another origin of the observed Raman peaks and the high frequencies of Raman peaks may be mostly under the influence of surface C-H bending modes. To resolve this uncertainty, we artificially increased the mass of terminating hydrogen atoms to 100 amu and repeated the whole of the calculations. This work helps us to determine whether the identified Raman peaks are associated with the C-C stretching modes, or C-H bending modes play the dominant role. This technique was previously used to remove the strong mixing modes in the Raman spectra of diamond nanoparticles terminated by hydrogen atoms [33]. By using this trick we suppress the role of C-H bending modes, which are more important in higher frequencies. We observe that the new results with heavy matrix reproduce the same three-active Raman modes that show again redshifts from the bulk diamond. Outcomes confirmed that the three redshifted high-frequency TO and LO optical modes along three orthogonal directions are originated from C-C stretching modes and are still Raman active. The corresponding peaks at between 1000-1100 cm$^{-1}$ show blueshift by changing the mass of the terminating hydrogen atoms. However, in overall, we do not see a significant change in the Raman spectra of DNWs below 1500 cm$^{-1}$, but it shows some shifts and changes in the intensities and number of the peaks. In another theoretical work, based on a local bond-polarization model, Raman scattering of DNWs has been investigated and it was shown that the highest optical Raman peaks of DNWs with diameters less than 2.50 nm experience redshifts with almost the same values as our prediction [11].

From 0.73 nm to 1.2 nm the LDA-predicted highest TO frequencies shifts slightly toward bulk diamond (from 1302 cm$^{-1}$ to 1313 cm$^{-1}$), hence we do not expect significant redshifts in large wires. The micro-Raman spectrum of DNWs with diameters between 60 and 90 nm growth in a carbon nanotube (CNT) shows a signature peak of the diamond crystal at 1332 cm$^{-1}$ with no redshift [3]. In previous Raman experiments of nanodiamond films and powders, some peaks at 1150 cm$^{-1}$ and 500 cm$^{-1}$ have been reported [34-35]. Previous theoretical works proposed that the observed peaks at 500 cm$^{-1}$ and 1150 cm$^{-1}$ do not belong to pure diamond nanoparticles and nano-films, but rather are originated from other sources, for example, defects in the sample [34]. DNWs in our study do not show any Raman peaks at 1150 cm$^{-1}$ or 500 cm$^{-1}$. The LDA calculated RBM peak at 509 cm$^{-1}$ for 0.73 nm wire cannot reminiscent the broad ~500 cm$^{-1}$ signals of Raman spectrum in nanodiamond powders, since RBM is highly size-dependent and LDA predicts a large redshift of about 200 cm$^{-1}$ in RBM frequencies in going from 0.73 nm to 1.2 nm DNWs.

Modifications of diamond phononic features in DNWs influence the related heat capacity. Previous *ab initio* calculations of electronic band structures reveal an insulating behavior for ultrathin H-terminated DNWs [13]. Therefore, we only concentrate on the contribution of the lattice dynamics to the heat capacity. The dependence of heat capacity on temperature *T* at constant volume is given by $C_V(T) = k_B \sum_\gamma \int_0^{+\infty} D(\omega) \left[\frac{\left(\frac{\hbar\omega}{k_B T}\right)^2 \exp\left(\frac{\hbar\omega}{k_B T}\right)}{\left(\exp\left(\frac{\hbar\omega}{k_B T}\right)-1\right)^2}\right] d\omega$, where D(ω) is the density of states at frequencies between ω and ω+dω and is calculated within DFPT for each branch of $\gamma$ [36]. The value of the function in the brackets of the integral becomes lower than 0.09 for $x = \frac{\hbar\omega}{k_B T} > 6$. Therefore, those modes with frequencies higher than the *lowest* optical mode $\omega_1$ have almost no contribution to the heat capacity at temperatures less than $T_1 = \frac{\hbar\omega_1}{6k_B}$ (~ 30-60 K for 0.73-1.7 nm [110] wire) and only acoustic modes can be excited. For two TA modes with $\omega \propto q^2$, $D(\omega) \propto \frac{1}{\sqrt{\omega}}$ and hence $C_V(T) \propto T^{0.5}$ at $T < T_0$. For the longitudinal and torsion modes $\omega \propto q$ and $D(\omega)$ = const, and consequently $C_V(T) \propto T$. So, one expects $C_V(T) \propto T^\alpha$, with $0.5 < \alpha < 1$ at lower temperatures. We applied the best-fitted power function to our data and found $C_V(T) \propto T^{0.76}$ at $T < T_0$ with the smallest root-mean-square error of 0.002 J/mol.K.

Several new states below 250 cm$^{-1}$ that are created due to confinement, now can be populated by thermal excitation at lower temperatures. Therefore, the C$_V$ curve of DNWs exceeds the bulk diamond at these low temperatures. At temperatures lower than 200 K, the C$_V$ of DNWs exceeds the corresponding value of the bulk diamond, and afterward, it decreases into the bulk value at about 310 K, as seen in Figure 6. At 320 K the C$_V$ is nearly 1% lower than the corresponding bulk value. Increasing the number of states that can be populated and redshifts in the optical modes reduce the heat capacity of DNWs from the bulk value at temperatures higher than 310 K.

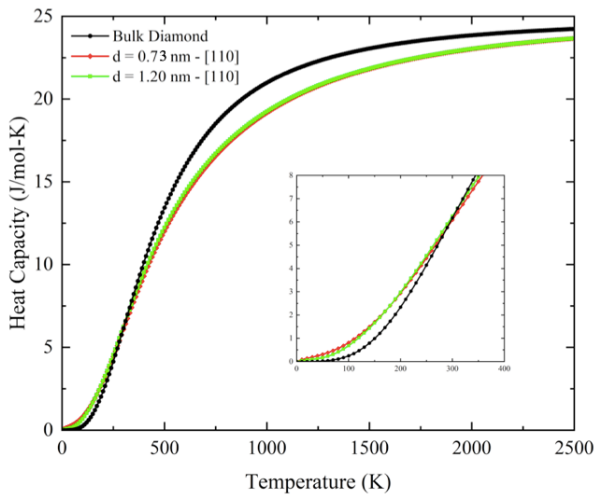

Figure 6. Calculated heat capacities (C$_V$, J/mol-K) of DNWs and bulk diamond. The low-temperature region is zoomed in for clarity.

## 3.3 Electronic Properties

### 3.3.1 H-terminated DNW

In this section, we first investigate the electronic properties of a fully H-terminated 1.2 nm DNW. Despite the general expectation from the quantum confinement effect and in agreement with the previous study [13], the calculated LDA-bandgap of the wire is 2.36 eV and significantly smaller than the gap of the bulk diamond (5.5 eV). The orbital-resolved band structure of this wire is shown in Figure 7(a). Due to confinement, the electronic states become less dispersive in comparison with the bulk diamond. The fully occupied states are extended in an energy interval of 21.24 eV, which is well comparable to the previously reported energy distribution of the occupied states in bulk diamond [15]. In the interval between about -22 to −10 eV below valence band, the band structure of the wire is dominated by the carbon $2s$ bonding interactions and represents the expected parabolic behavior with $\mathbf{k}$. The higher occupied bands, based on the σ-bonding of carbon $2p$ orbitals, start in about -8 eV at Γ-point from the top of the valence band. Contrary to the parabolic curvature of $2s$ bonding and in a way similar to the bulk diamond, the bands resulting from $2p_\sigma$ bonding interaction show a clear parabolic decrease with moving away from Γ-point. The bandwidth of the valence band is 3.5 eV and its maximum happens at Γ-point. A significant contribution of the hydrogen atoms to the conduction band is clearly seen in pDOS presented in Figure 7(a) and manifest the C-H antibonding character of the empty states.

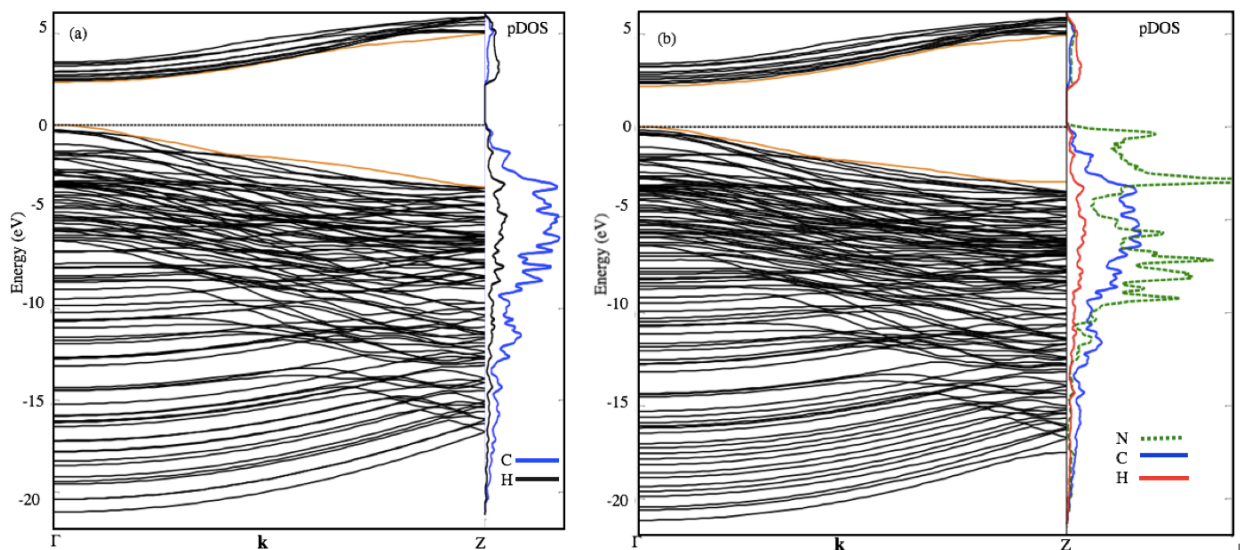

Figure 7. Band structures of 1.2 nm of (a) H-DNW and (b) $NH_2$-functionalized DNW. Normalized PDOS (to the number of the corresponding atom in the unit cell) of the carbon $2p$, hydrogen $1s$ and nitrogen $2p$ orbitals are shown to figure out the contribution of the corresponding atomic orbitals to the band structure. Valence band maximum in both figures are set to zero.

In Figure 9(a), we show isosurface plots of the valence and conduction bands of the wire at Γ-point. Just comparable to the highest occupied molecular orbital (HOMO) of diamondoids, the valence orbitals are nearly localized on C-C bonds. On the other hand, 1s hydrogen orbitals characterize the conduction band, as also concluded from Figure 7(a). Isosurface plots of the conduction band in Figure 9(a) indicate that this band is delocalized over the hydrogen atoms, and is reminiscent of the lowest molecular unoccupied orbital (LUMO) of diamondoids [37-38]. Importantly, this band falls below the conduction band of the bulk diamond and makes the bandgap smaller in DNWs.

*3.3.2 NH₂-functionalized DNW*

To see the electronic response of DNWs when functionalized with an amine-group, we replace one of the hydrogen atoms on the wire surface with an NH$_2$ radical. Previous scanning tunneling microscope (STM) image identified two distinct (111) and (100) facets for Si-NW align along [110] direction, see Figure 8(a) [39-40]. It is probable that the contribution of the NH$_2$ group to the electronic properties of DNWs depends on the corresponding atomic patterns on the given facet. Consequently, we have constructed two distinct calculations of NH$_2$-functionalized DNWs in (111) and (100) facets, according to Figure 8(a). We compared the binding energy ($\Delta E_B$) of NH$_2$ on each facet by using $\Delta E_B = E(DNW - 1H) + E(NH_2) - E(NH_2@DNW)$ equation. Here $E(NH_2)$ and $E(DNW - 1H)$ are respectively the total energy of an isolated NH$_2$ radical and an isolated fully H-terminated 1.2 nm DNW when one hydrogen atom is removed and a dangling bond is left behind. $E(NH_2@DNW)$ is the total energy of the corresponding wire, when the dangling bond is passivated by an NH$_2$ radical. At each step, each structure was fully relaxed separately. In calculating binding energy, we care about the spin asymmetry which may induce a magnetic moment in the ground state. Especially, the spin polarization calculations are employed to investigate the total energy ground state of the wire with one removed hydrogen atom ($DNW - 1H$). The number of the total electrons in a unit cell of bulk diamond and in the supercell of the fully DNW in our study are even. Spin-polarized calculations show non-magnetic ground state for the fully hydrogen-passivated DNW. However, the situation may be changed if the DNW is not completely saturated by hydrogen atoms. By removing one H atom from the surface of DNW, it is left with an odd number of electrons in the unit cell. This situation requires spin-polarized calculations. Spin polarized calculations lower the ground-state total energy of the system by 161 meV, when compared with spin unpolarized calculations. DFT predicts a total magnetic moment of M=1 μ$_B$ for this structure, where μ$_B$ is Bohr magnetic moment. Most of the 1 μ$_B$ of magnetization (~80%), previously carried by the hydrogen atoms, is now induced on the unsaturated carbon atom. In Figure 8(c) we pictorially display the difference between the spin-up and spin-down electron densities *(n↑-n↓)* resulting in a net magnetic moment in the wire. A comparison between the total electronic density of states of the majority and minority spins and partial density of states (pDOS) of unsaturated carbon atom in Figure 8(b) shows that the unpaired $p_x$-orbital of unsaturated carbon atom in the perpendicular direction to the nanowire axis determines the magnetic response of the structure.

To continue, we name (111)- and (100)-facets NH$_2$-functionalized wires DNW1 and DNW0, respectively. The periodic lattice parameter "a" of DNW1 after full relaxation increases slightly from 2.529 to 2.531 Å, while "a" does not change for DNW0. Phonon band structure shows no negative value. DNW1 and DNW0 exhibit noticeable binding energies of 3.93 eV and 3.73 eV, respectively, hence functionalizing on (111) facet is more favored than on (100) facet by 0.2 eV. Previously, it was shown that amine groups reduce the bandgap of smaller diamondoids and impose a new level inside the gap [40, 41]. This subject has also been confirmed by the experiments [42]. Here also NH$_2$ reduces the bandgap from 2.36 eV to 2.18 eV. That the amine radical bound to which of facet does not influence bandgap and (almost) valence bandwidths. In both cases, valence bandwidth reduces from 3.40 eV to about 3 eV. Conduction bandwidths of DNW1 and DNW0 are influenced slightly by the NH$_2$ and change about 0.13 eV and -0.21 eV, respectively.

Figure 8. (a) Facets on a DNW with axis along [110] direction, as identified by STM imaging in [39]. (b). PDOS of spin-up (↑) and spin-down (↓) of non-passivated carbon 2p orbitals. A clear polarization in xoy plane perpendicular to the wire axis is observed which results in forming a local magnetic moment. (c) Pictorial illustration of the difference between the spin-up and spin-down electron densities in green *(n↑-n↓)* on the base plane of the 1.2 nm DNW shows a net local magnetic moment (M=1 $\mu_B$) is produced as a result of removing a hydrogen atom from the surface. The corresponding magnetic moment is localized on the non-pasivated surface carbon atom. (d) PDOS of thoes atoms contributed to C-N and N-H bonding including nitrogen 2p orbitals, two hydrogen atoms 1s orbitals, and one carbon 2p orbital. The valence band maximum is set to zero.

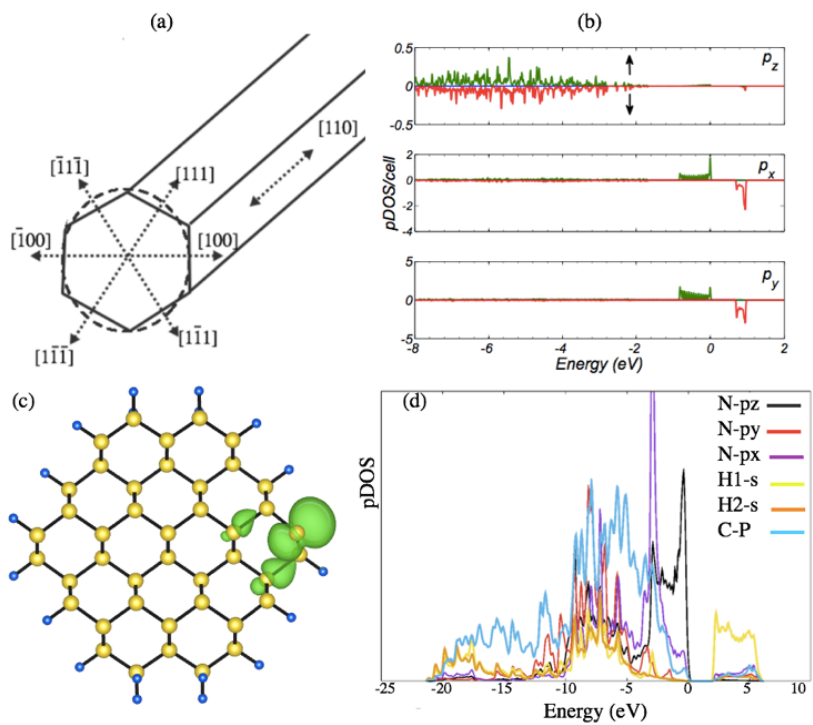

From fully H-terminated DNW to NH$_2$-functionalized DNW, the valence isosurface moves somewhat toward the NH$_2$ group, see Figure 9(b). Likewise, orbital-resolved DOS in Figure 7(b) exhibits a significant contribution of the nitrogen atom to the top of the valence band. We are interested to figure out the role of the nitrogen nonbonding electron pairs on the valence band. pDOS of the nitrogen 2p orbitals along with the carbon atom involved in C-N bond and 1s orbital of those hydrogen atoms involved in N-H bonds are shown in Figure 8(d). From this figure, one concludes that the main contribution of nitrogen non-bonding pair electrons may be at the top of the valence band, where the carbon 2p orbitals and the hydrogen atoms 1s orbitals are almost absent. According to this figure, the dominant contributions of the C-N and N-H bonds are in lower energy windows between -10 eV and -2.5 eV, respectively. LUMO still shows a diffused character but over those of hydrogen atoms that are next to NH$_2$-radical.

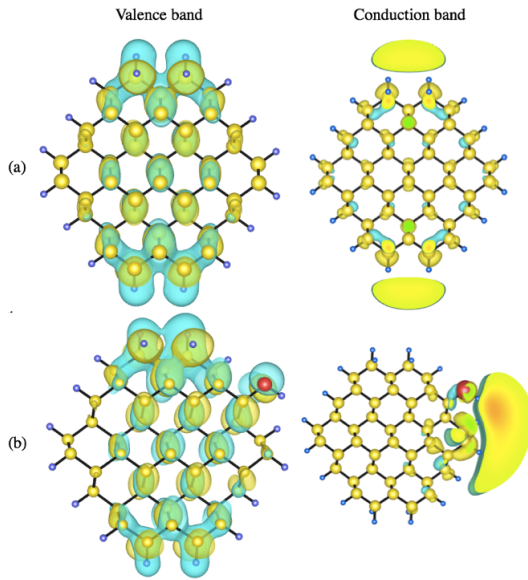

Figure 9. (color online) Isosurfaces plot (iso value=0.0005 a. u.) of the valence and conduction bands obtained at the Γ-point for 1.2 nm (a) H-DNW (b) NH$_2$-functionalized DNW.

## 4. Conclusions

In conclusion, we have performed a detailed discussion on both the lattice dynamics and the electronic features of DNWs. We have classified two regions of vibrational modes for ultrathin DNWs: one is low-frequency characterized with confined modes and the other is the high-frequency region with significant contributions of hydrogen atoms. Both types show a size-dependent trend, but with an opposite behavior: blueshift in low frequencies which are mostly influenced by the confinement effect, redshift in high-frequency which are mostly influenced by the surface termination effect. By artificially increasing the mass of hydrogen atoms, we have concluded that the highest-frequency optical modes, below 1500 cm$^{-1}$, are mainly originated from C-C stretching and bending modes. We have shown that the heat capacity of DNWs is larger than bulk diamond at room temperature, however, with increasing temperature it falls below the bulk value. We have also investigated the electronic features of DNW. DOS orbital analysis shows that the conduction band, even when the wire is functionalized by an amine group, is a surface state which its energy falls inside the bulk diamond bandgap, and makes band gap of DNWs smaller than bulk diamond. Our results show that the binding energy of $NH_2$-radical to (111)-facet is larger than to (100)-facet. In both cases, $NH_2$ reduces the band gap and this is independent of which facet is functionalized by $NH_2$ radical.

## ACKNOWLEDGMENTS

The authors gratefully acknowledge the computational assistance provided by Dr. Salman Noorazar at the Amirkabir University of Technology.

**Refrences**